\documentclass[10pt,conference]{IEEEtran}

\usepackage{times}
\usepackage{latexsym}
\usepackage[T1]{fontenc}
\usepackage[utf8]{inputenc}
\usepackage{microtype}
\usepackage{inconsolata}
\usepackage{cite}
\usepackage{graphicx}
\usepackage[ruled,vlined,lined,commentsnumbered]{algorithm2e}
\usepackage{amsmath,amsfonts}
\usepackage{amssymb}
\usepackage{array}
\usepackage{booktabs}
\usepackage[skip=1pt,labelfont=bf]{caption}
\usepackage{calligra}
\usepackage{color, colortbl}
\usepackage{courier}
\usepackage{csvsimple}
\usepackage{enumitem}
\usepackage{fancybox}
\usepackage{listings}
\usepackage{longtable}
\usepackage{lscape}
\usepackage{makecell}
\usepackage{marvosym}
\usepackage{moreverb}
\usepackage{multicol}
\usepackage{multirow}
\usepackage{pifont}
\usepackage{rotating}
\usepackage{setspace}
\usepackage{caption}
\usepackage{subfigure}
\usepackage{parcolumns}
\usepackage[most]{tcolorbox}
\usepackage{threeparttable}
\usepackage{tikz}
\usepackage[normalem]{ulem}
\usepackage[hyphens]{url}
\usepackage{soul}
\usepackage{wasysym}
\usepackage{svg}
\usepackage{textcomp}
\usepackage{xcolor}
\usepackage{wrapfig}
\usepackage{pdflscape}
\usepackage{hyperref}
\usepackage{tikz}

\usepackage{amsmath}
\usepackage{array}
\usepackage{balance}
\colorlet{punct}{red!60!black}
\definecolor{background}{HTML}{EEEEEE}
\definecolor{delim}{RGB}{20,105,176}
\definecolor{yellow}{RGB}{254, 254, 0}
\definecolor{lightblue}{RGB}{0, 254, 254}
\colorlet{numb}{magenta!60!black}

\lstdefinestyle{sol}{
    language=Python,
    basicstyle=\ttfamily\small,
    keywordstyle=\color{blue!70},
    commentstyle=\color{green!50!black},
    stringstyle=\color{red!70},
    numbers=left,
    numberstyle=\tiny\color{gray},
    stepnumber=1,
    showspaces=false,
    showstringspaces=false,
    breaklines=true,
    frame=single,
    tabsize=4,
    morekeywords={function, address, external, returns, internal, assertEq},
    escapeinside=`` 
}

\lstdefinelanguage{json}{
    basicstyle=\normalfont\ttfamily,
    numbers=left,
    numberstyle=\scriptsize,
    stepnumber=1,
    numbersep=8pt,
    showstringspaces=false,
    breaklines=true,
    frame=lines,
    backgroundcolor=\color{background},
    literate=
     *{0}{{{\color{numb}0}}}{1}
      {1}{{{\color{numb}1}}}{1}
      {2}{{{\color{numb}2}}}{1}
      {3}{{{\color{numb}3}}}{1}
      {4}{{{\color{numb}4}}}{1}
      {5}{{{\color{numb}5}}}{1}
      {6}{{{\color{numb}6}}}{1}
      {7}{{{\color{numb}7}}}{1}
      {8}{{{\color{numb}8}}}{1}
      {9}{{{\color{numb}9}}}{1}
      {:}{{{\color{punct}{:}}}}{1}
      {,}{{{\color{punct}{,}}}}{1}
      {\{}{{{\color{delim}{\{}}}}{1}
      {\}}{{{\color{delim}{\}}}}}{1}
      {[}{{{\color{delim}{[}}}}{1}
      {]}{{{\color{delim}{]}}}}{1},
}

\newcommand{\mytitle}{PrefGen\xspace}
\newcommand{\datasetname}{PrefGen\xspace}

\newcommand{\intuition}[1]{
\begin{tcolorbox}[colback=white,boxrule=1pt,top=0pt,bottom=0pt,left=1pt,right=2pt,top=2pt,bottom=2pt]
\em #1
\end{tcolorbox}
}

\begin{document}
\title{
\mytitle: A Preference-Driven Methodology for Secure Yet Gas-Efficient Smart Contract Generation
}

\author{
Zhiyuan Peng$^{1}$\IEEEauthorrefmark{2}
\quad 
Xin Yin$^{2}$\IEEEauthorrefmark{2}
\quad 
Zijie Zhou$^{3}$
\quad
Chenhao Ying$^{1}$\IEEEauthorrefmark{1} 
\quad 
Chao Ni$^{2}$\IEEEauthorrefmark{1} 
\quad 
Yuan Luo$^{1}$\IEEEauthorrefmark{1} 
\quad 
\\
\IEEEauthorblockA{
\textit{$^{1}$Shanghai Jiao Tong University}, \{pzy2000, yingchenhao, yuanluo\}@sjtu.edu.cn}
\IEEEauthorblockA{
\textit{$^{2}$The State Key Laboratory of Blockchain and Data Security, Zhejiang University}, \{xyin, chaoni\}@zju.edu.cn}
\IEEEauthorblockA{
\textit{$^{3}$China University of Petroleum (Beijing)}, zhouzijie@student.cup.edu.cn}
}

\maketitle

\begingroup
\renewcommand\thefootnote{\IEEEauthorrefmark{2}}
\footnotetext{
Equal contribution.}
\renewcommand\thefootnote{\IEEEauthorrefmark{1}}
\footnotetext{
Corresponding authors.\\
Chao Ni is also with Hangzhou High-Tech Zone (Binjiang) Institute of Blockchain and Data Security.}
\endgroup

\begin{abstract}
    While Large Language Models (LLMs) have demonstrated remarkable progress in generating functionally correct Solidity code, 
    they continue to face critical challenges in producing gas-efficient and secure code, which are critical requirements for real-world smart contract deployment.
    Although recent advances leverage Supervised Fine-Tuning (SFT) and Direct Preference Optimization (DPO) for code preference alignment, existing approaches treat functional correctness, gas optimization, and security as independent objectives, resulting in contracts that may achieve operational soundness but suffer from prohibitive execution costs or dangerous vulnerabilities.
    To address these limitations, we propose \textbf{\mytitle}, a novel framework that extends standard DPO beyond human preferences to incorporate quantifiable blockchain-specific metrics, enabling holistic multi-objective optimization specifically tailored for smart contract generation.
    Our framework introduces a comprehensive evaluation methodology with four complementary metrics: Pass@k (functional correctness), Compile@k (syntactic correctness), 
    Gas@k (gas efficiency), and Secure@k (security assessment), providing rigorous multi-dimensional contract evaluation.
    Through extensive experimentation, we demonstrate that \mytitle significantly outperforms existing approaches across all critical dimensions,
    achieving 66.7\% Pass@5, 58.9\% Gas@5, and 62.5\% Secure@5, while generating production-ready smart contracts that are functionally correct, cost-efficient, and secure.
\end{abstract}

\begin{IEEEkeywords}
Smart Contract Generation, Supervised Fine-Tuning, Direct Preference Optimization
\end{IEEEkeywords}

\section{Introduction}

Large Language Models (LLMs) have demonstrated remarkable capabilities in generating functionally correct Solidity code, 
as evidenced by recent benchmarks~\cite{benchmark2024sol,peng2025soleval}. 
However, while functional correctness represents a fundamental requirement, human preferences for smart contracts extend far beyond mere operational soundness.
In practice, developers and users demand contracts that simultaneously achieve security-critical and gas efficiency attributes that directly impact deployment costs and vulnerability exposure in real-world blockchain environments.

This multi-dimensional preference presents a significant challenge for current LLM-based smart contract generation approaches. 
Security vulnerabilities in LLM-generated smart contracts pose substantial risks for blockchain deployment, as Solidity's contract-oriented nature makes it susceptible to various exploits, including reentrancy attacks~\cite{atzei2017survey}, integer overflows~\cite{durieux2020empirical}, and access control vulnerabilities~\cite{ghaleb2023achecker}. 
The severity of these risks was starkly demonstrated in May 2025 when Cetus Protocol, the largest decentralized exchange on the Sui blockchain, 
suffered a devastating exploit that drained over \$260 million through smart contract vulnerabilities involving manipulated price 
curves and flawed reserve calculations\footnote{\url{https://www.theblock.co/post/355906/cetus-poses-community-vote-to-possibly-return-100-of-funds-to-users-affected-by-223-million-exploit}}. 
Previous work~\cite{peng2025soleval} also reveals that LLMs performing well on Pass@k may produce code with elevated vulnerability rates,  indicating a fundamental misalignment between model optimization objectives and human security preferences.

Equally critical is the gap in gas efficiency optimization, which represents another dimension of human preference that current LLMs largely overlook.
Gas efficiency is a paramount concern for Ethereum deployment, where execution costs directly impact user adoption and economic viability~\cite{ghaleb2023achecker,peng2025mulchain}. 
While models like GPT-4~\cite{openai_access_gpt4} and DeepSeek-R1~\cite{deepseekr1} excel at producing functionally correct contracts, they typically neglect gas consumption optimization, resulting in contracts that meet functional requirements but violate user preferences for cost-effectiveness~\cite{chen2025chatgpt,peng2025soleval}. 
The fundamental issue lies in the misalignment between current optimization objectives and genuine human preferences for smart contracts.
Optimizing for one dimension often sacrifices others~\cite{chen2025chatgpt,momeni2019machine}, creating a critical trade-off that is particularly problematic in decentralized finance (DeFi) applications, where high gas costs can undermine economic feasibility~\cite{gudgeon2020decentralized,werner2022sok}. 
Although recent advances leverage Supervised Fine-Tuning (SFT) and Direct Preference Optimization (DPO) for code preference alignment,
existing approaches address correctness, efficiency, and security in isolation~\cite{peng2025soleval,momeni2019machine}, failing to capture the holistic preference optimization required for production-ready smart contracts that satisfy real-world user requirements~\cite{chen2025chatgpt}.

To address these limitations, we introduce \textbf{\mytitle}, 
a novel framework that extends standard DPO beyond subjective preference pairs to incorporate quantifiable metrics that reflect genuine user priorities (i.e., gas-efficiency and security), enabling comprehensive preference-driven optimization tailored for smart contract development.
Unlike existing methods that treat these objectives independently, \mytitle learns to favor solutions that minimize gas consumption while maintaining correctness and avoiding vulnerabilities through a custom DPO implementation that incorporates domain-specific reward signals. 
Specifically, gas efficiency rewards encourage lower consumption and security rewards penalize vulnerabilities, enabling simultaneous multi-objective optimization that aligns with real-world deployment preferences.

\mytitle employs a systematic approach to construct preference datasets by generating multiple candidate implementations for each functional requirement and evaluating them across three critical dimensions that capture real-world deployment requirements: functional correctness (i.e., Pass@k), gas efficiency (i.e., Gas@k), and security (i.e., Secure@k). 
This multi-dimensional evaluation framework enables the creation of preference pairs that guide the model toward generating production-ready smart contracts that simultaneously satisfy functional, gas-efficiency, and security constraints, addressing the comprehensive spectrum of developer and user preferences in blockchain environments.

Our comprehensive evaluation demonstrates that \mytitle significantly outperforms existing approaches across all three critical dimensions, bridging the gap between academic benchmarks and production deployment requirements. 
Compared to the pretrained Qwen-7B baseline (16.7\% Pass@5, 0.0\% Gas@5, 11.8\% Secure@5), 
\mytitle achieves substantial improvements: 66.7\% Pass@5, 58.9\% Gas@5, and 62.5\% Secure@5.
Beyond these metrics, practical deployment scenarios demonstrate real-world impact: 
\mytitle generates ERC-20 contracts with 12\% reduced gas costs compared to baseline models while completely eliminating critical security vulnerabilities such as reentrancy attacks, exemplifying the framework's ability to deliver contracts that meet the stringent requirements of production blockchain environments.

In summary, our contributions are:

\begin{itemize}[leftmargin=*]
    \item \textbf{Comprehensive Multi-Dimensional Evaluation:} 
    We establish a systematic evaluation framework with four complementary metrics to provide a rigorous multi-dimensional assessment of generated smart contracts. 
    Our extensive evaluation on smart contracts across 16 SOTA LLMs reveals that even top-performing models struggle with gas efficiency and security, demonstrating substantial room for improvement.

    \item \textbf{Novel Preference-Driven Framework:} We introduce \mytitle, a novel framework that extends DPO beyond subjective  preference pairs to incorporate quantifiable blockchain-specific metrics (e.g., gas-efficiency and security).
    
    \item \textbf{Production-Ready Smart Contract Generation:} \mytitle achieves breakthrough performance improvements: 66.7\% Pass@5, 58.9\% Gas@5, and 62.5\% Secure@5 on Qwen-7B, with practical deployment benefits including 12\% reduced gas costs in ERC-20 contracts and complete elimination of critical security vulnerabilities such as reentrancy attacks.
    We open-source our dataset and evaluation framework at~\cite{replication}.

\end{itemize}
\label{sec:introduction}

\section{Background and Motivation}

\subsection{Background}
\subsubsection{Security Vulnerabilities in Smart Contracts}
Smart contracts operate in an immutable and trustless environment where security vulnerabilities can lead to catastrophic financial losses. 
The critical nature of smart contract security was demonstrated in May 2025 when Cetus Protocol suffered a devastating exploit that drained over \$260 million through manipulated price curves and flawed reserve calculations~\cite{immutable1}. 
This incident exemplifies how implementation flaws can cascade into systemic failures, where billions of dollars in assets remain at risk.
The immutable nature of blockchain deployments distinguishes smart contract security from traditional software development. 
Unlike conventional applications, smart contracts cannot be modified once deployed~\cite{defi_attack_survey,smart_contract_security_survey}, making prevention the only viable defense strategy. 
This constraint has resulted in cumulative losses exceeding billions of dollars~\cite{defi_financial_losses_2022,blockchain_security_incidents,smart_contract_attack_statistics}, highlighting the paramount importance of generating inherently secure code.

Common vulnerability patterns include \textit{reentrancy attacks} that exploit external calls to recursively drain funds~\cite{atzei2017survey}, 
\textit{integer overflow/underflow vulnerabilities} from unchecked arithmetic operations~\cite{durieux2020empirical}, 
and \textit{access control vulnerabilities} from improper permission management~\cite{ghaleb2023achecker}. 
Traditional static analysis tools like Slither~\cite{feist2019slither} and SolTG~\cite{britikov2024soltg} systematically study these vulnerability classes.
Recent advances leverage LLMs for smart contract security analysis~\cite{tse24gassmell}, demonstrating capabilities in identifying complex vulnerability patterns. 
However, these approaches focus on post-development analysis rather than prevention during generation. 
Critically, empirical studies reveal a fundamental misalignment: models achieving high functional correctness may simultaneously generate code with elevated vulnerability rates~\cite{peng2025soleval}, indicating that traditional training paradigms inadequately address security considerations.

Production deployment demands simultaneous optimization across security, gas efficiency, and maintainability dimensions~\cite{defi_security_analysis,smart_contract_vulnerabilities_ethereum}. 
Existing LLM-based approaches treat these objectives independently, failing to capture the holistic optimization required for production-ready contracts. 
This necessitates training frameworks that integrate security considerations directly into the code generation process.

\subsubsection{Gas Efficiency Challenges}
While LLMs have demonstrated proficiency in generating functionally correct Solidity code~\cite{benchmark2024sol, peng2025soleval}, they consistently produce contracts with suboptimal gas consumption. 
This limitation poses barriers to practical deployment, where gas costs directly impact user adoption and economic viability.

Consider the deployment of machine learning models on Ethereum: uploading weights and test data can require $7.3 \times 10^{7}$ and $2.8 \times 10^{9}$ gas units respectively, translating to deployment costs of approximately 0.054 ETH (\$13.40) and 0.2949 ETH (\$737.22)~\cite{sham2025generation}. These figures underscore the critical need for gas-optimized code generation in blockchain applications.

The SolEval benchmark~\cite{peng2025soleval} incorporates gas efficiency as an evaluation metric for repository-level Solidity generation. However, while it measures gas consumption, it lacks mechanisms for optimization. Results reveal a persistent trade-off: models achieving high functional correctness often generate contracts with prohibitive gas costs, highlighting the need for integrated optimization frameworks.

\subsubsection{SFT and DPO}
Supervised Fine-Tuning (SFT) and Direct Preference Optimization (DPO) represent two complementary paradigms for aligning LLMs with specific domain requirements and human preferences~\cite{rafailov2023direct,zhang2024instruction}. 
While traditional pre-training optimizes models for next-token prediction on vast text corpora, these post-training techniques enable targeted adaptation for specialized tasks and quality criteria.

\textit{Supervised Fine-Tuning} adapts pre-trained models to specific domains through supervised learning on curated instruction-response pairs~\cite{zhang2024instruction}. In the context of code generation, SFT trains models to map natural language specifications to syntactically correct and functionally accurate implementations. This process fundamentally shifts the model's objective from general language modeling to domain-specific code synthesis, establishing a foundation for functional correctness that is essential for smart contract development.

\textit{Direct Preference Optimization} addresses limitations of traditional Reinforcement Learning from Human Feedback by directly optimizing language models using preference data~\cite{rafailov2023direct}. 
DPO treats the language model itself as an implicit reward model, enabling more stable and efficient preference learning. 
This paradigm is particularly valuable for code generation, where multiple valid implementations exist but differ significantly in quality dimensions such as efficiency, security, and maintainability~\cite{zhang2024codedpo}.
For Solidity smart contract generation, 
these techniques address fundamental challenges that pure pre-training cannot resolve. 
First, smart contracts operate under unique constraints (e.g., gas costs, security vulnerabilities) and blockchain-specific semantics that are underrepresented in general programming corpora~\cite{chen2025chatgpt,peng2025soleval}. 
SFT enables models to learn these domain-specific patterns through targeted training on Solidity implementations. 
Second, functional correctness alone is insufficient for production deployment~\cite{zhang2024codedpo}; 
contracts must simultaneously optimize for gas efficiency and security~\cite{chen2017under,kong2022characterizing}. 
DPO provides a principled framework for learning these multi-objective trade-offs through preference learning, 
where models learn to favor implementations that balance correctness, efficiency, and security.

Recent work has demonstrated the effectiveness of combining SFT and DPO for code generation tasks~\cite{zhang2024codedpo,yin2025learning}, showing that this two-phase approach enables models to first establish functional competence through SFT, then refine quality through preference optimization. For smart contract generation, this progression is particularly critical: establishing basic Solidity syntax and semantics through SFT creates a foundation for subsequent optimization of blockchain-specific objectives through DPO.

\subsection{Preliminary Experiment}
We first conduct a preliminary experiment to evaluate the performance of LLMs on Solidity code generation. 
We evaluate 16 state-of-the-art LLMs on SolEval~\cite{peng2025soleval} with Task@k.

\textit{Results reveal significant limitations in current LLMs for smart contract generation.} 
While the best-performing model (DeepSeek-V3) achieves only 24.99\% Pass@5, 
indicating that even state-of-the-art LLMs struggle with functional correctness. 
More critically, gas efficiency remains severely compromised: the highest Gas@5 score is merely 7.13\% (DeepSeek-V3), 
meaning over 92\% of functionally correct contracts exhibit suboptimal gas consumption. Security vulnerabilities present similar challenges, 
with detection rates ranging from 5.28\% to 19.29\% across models, 
indicating significant room for improvement in both efficiency and security.

These findings demonstrate a fundamental trade-off: models achieving higher functional correctness 
(e.g., DeepSeek-V3 with 24.99\% Pass@5) do not necessarily excel in gas efficiency (7.13\% Gas@5) or security (19.29\% Secure@5). 
Conversely, smaller models like GPT-4o-mini show better gas efficiency (2.42\% Gas@5) but lower functional correctness (12.37\% Pass@5). 
The results are shown in Table~\ref{tab:rq1}.

\subsection{Motivation}
To address these fundamental limitations, we propose \textbf{\mytitle}, a framework that integrates multi-objective optimization into the training process itself. 
Unlike existing approaches that treat gas efficiency and security as post-hoc evaluation metrics, \mytitle incorporates these objectives directly into the learning process through extended DPO.

Our framework addresses three key gaps in current research. 
Rather than optimizing objectives independently, \mytitle employs preference learning to balance functional correctness, gas efficiency, and security simultaneously, representing a significant advancement in integrated optimization approaches. 
Furthermore, we extend standard DPO with quantifiable blockchain-specific metrics, enabling optimization tailored for smart contract deployment requirements through domain-specific rewards. 
Finally, \mytitle generates contracts that meet real-world deployment criteria, effectively addressing the gap between academic benchmarks and production requirements with a practical deployment focus.

\begin{figure*}[htbp!] 
    \centering
    \includegraphics[width=0.85\linewidth]{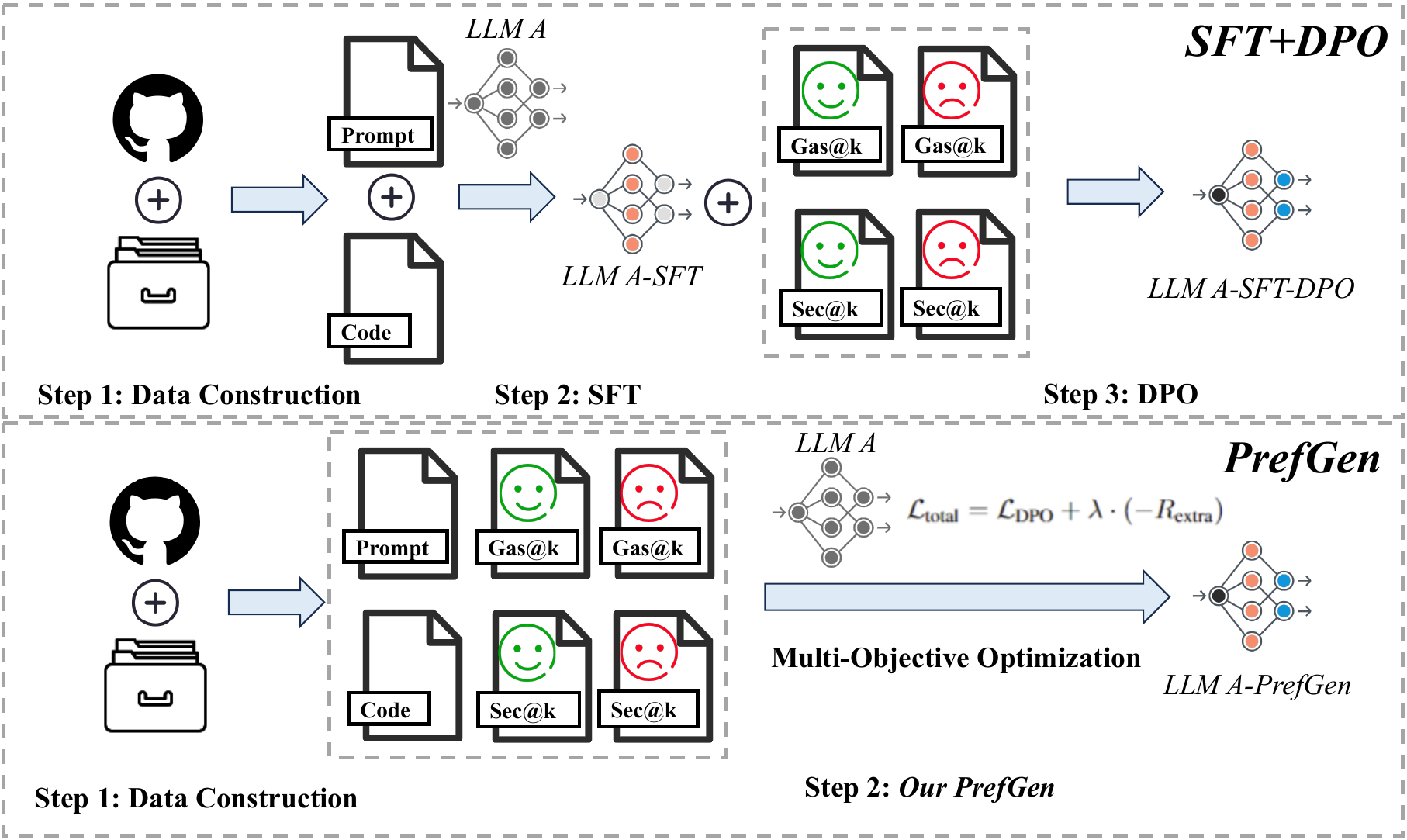}
    \caption{Overview of the \mytitle framework v.s SFT + DPO. }
    \label{fig:overview}
\end{figure*}
\label{sec:motivation}

\section{Approach}

This section presents \textit{\mytitle}, a framework that leverages multi-objective optimization to generate functionally correct, gas-efficient, and secure Solidity smart contracts.

\subsection{Framework Overview}

As shown in Figure~\ref{fig:overview}, we first construct a two-phase training pipeline that progressively refines code generation capabilities. 
The \textit{SFT phase} establishes functional correctness by training on verified Solidity contracts with comprehensive test suites. 
The \textit{DPO phase} then optimizes for gas efficiency and security through preference learning, 
where the model learns to favor solutions that minimize gas consumption while maintaining correctness and avoiding vulnerabilities.

Then, we extend standard DPO beyond human preferences to incorporate quantifiable metrics (i.e., gas consumption and security vulnerabilities), 
enabling multi-objective optimization tailored for smart contract development.

\subsection{Data Quality Ranking Algorithm}

To ensure high-quality training data across both SFT and DPO phases, we employ a PageRank-inspired algorithm~\cite{page1999pagerank} that iteratively ranks code snippets and test cases based on mutual validation performance. This ranking mechanism is essential for identifying the most reliable training samples throughout our framework.

We initialize each code snippet and test case with a unit score. Through $M = 10$ iterations, scores are updated based on cross-validation performance:

\begin{align}
\text{Score}^m(c) &= (1 - d) \times \text{Score}^{m-1}(c) \nonumber \\
&\quad + d \times \sum_{t} \text{Score}^{m-1}(t) \times \text{Link}(t, c)
\end{align}

\begin{align}
\text{Score}^m(t) &= (1 - d) \times \text{Score}^{m-1}(t) \nonumber \\
&\quad + d \times \sum_{c} \text{Score}^{m-1}(c) \times \text{Link}(c, t)
\end{align}

where $d$ is the damping factor and $\text{Link}(t, c)$ indicates whether code snippet $c$ passes test case $t$. 
Following established PageRank practices~\cite{page1999pagerank,zhang2024codedpo}, 
we use $d = 0.85$. Our ablation analysis demonstrates this choice: at $d = 0.50$, 
the model achieves 63.1\% Pass@5, 52.4\% Gas@5, and 56.0\% Secure@5; 
increasing to $d = 0.65$ yields 64.8\%, 54.3\%, 58.2\%; at $d = 0.75$ produces 66.0\%, 58.0\%, 61.3\%; 
the canonical $d = 0.85$ peaks at 66.7\%, 58.9\%, 62.5\%; 
while $d = 0.95$ scores decline to 65.4\%, 57.7\%, 60.8\%. 
This iterative refinement converges to rankings that accurately reflect code quality based on correctness validation. 
Higher-ranked implementations are prioritized during training, 
ensuring the model learns from verified, high-quality exemplars.

\subsection{Supervised Fine-Tuning (SFT)}

The SFT phase trains an LLM to generate functionally correct Solidity code from natural language specifications.

\subsubsection{Dataset Construction}

We carefully curate a dataset from a large-scale real-world Solidity benchmark~\cite{peng2025soleval}, where each code sample is paired with several comprehensive test cases defining expected behavior. The data quality ranking algorithm described above is applied to select the highest-quality code-test pairs for training.

\subsubsection{Training Objective}

The model is fine-tuned using cross-entropy loss to minimize the divergence between predicted and ground-truth code:
\begin{equation}
\mathcal{L}_{\text{SFT}} = -\sum_{i=1}^{N} \log P(y_i | x_i, \theta)
\end{equation}
where $x_i$ represents functional specifications, $y_i$ the corresponding Solidity implementation, and $\theta$ the model parameters. This phase optimizes the Pass@k metric, measuring the proportion of generated solutions that satisfy functional requirements.

\begin{figure*}[htbp!]
    \centering
    \includegraphics[width=.85\linewidth]{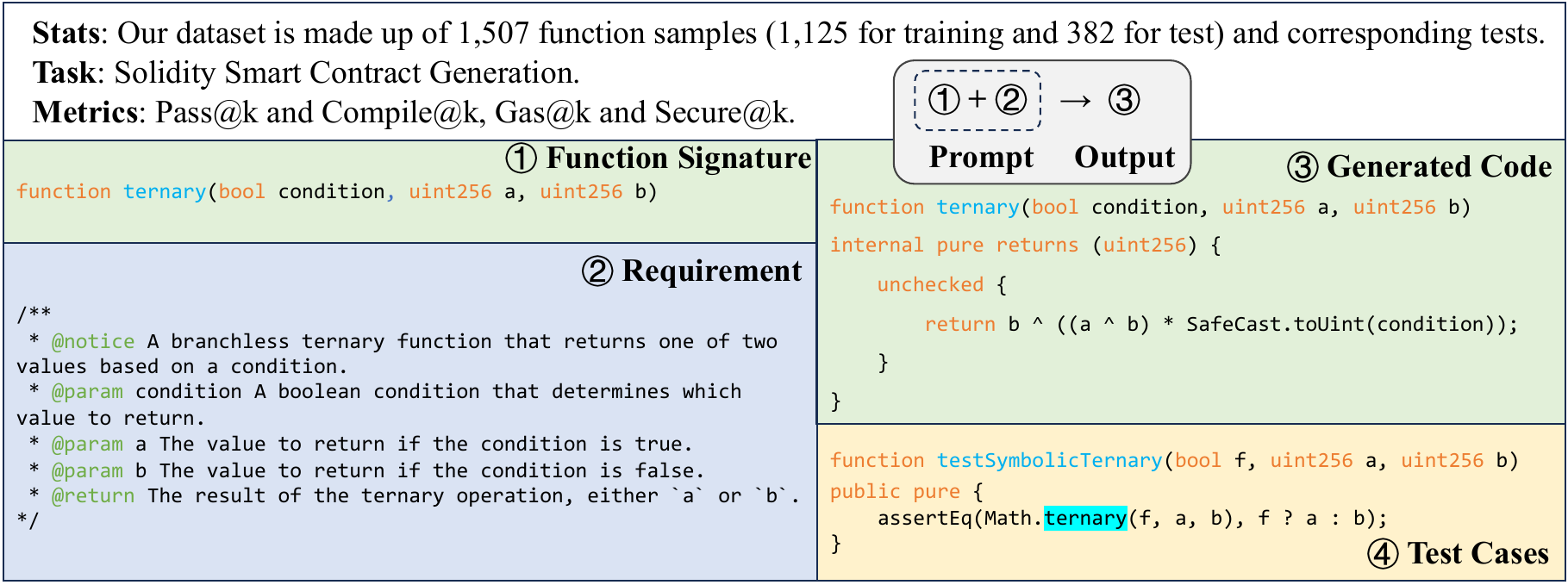}
    \caption{Prompt construction for \mytitle.}
    \label{fig:prompt_construction}
\end{figure*}

\subsection{Direct Preference Optimization (DPO)}

The DPO phase refines the SFT model to generate gas-efficient and secure code through preference learning.

\subsubsection{Preference Dataset Construction}

For each functional specification, we generate multiple candidate implementations and evaluate them on Task@k metrics.
Preference pairs are constructed by ranking candidates using our data quality ranking algorithm, 
where solutions with higher Task@k are preferred. 
The details of the metrics are provided in Section~\ref{sec:eval_metric}.

\subsubsection{Multi-Objective Optimization}

We extend standard DPO to incorporate domain-specific objectives beyond human preferences. Our approach introduces additional reward terms that capture gas efficiency and security:

\paragraph{Gas Efficiency Reward:}
\begin{equation}
R_g = -(gas_{\text{chosen}} - gas_{\text{rejected}})
\end{equation}
where $gas_{\text{chosen}}$ and $gas_{\text{rejected}}$ represent the gas consumption of the chosen 
and rejected solutions, respectively. 
This formulation encourages the model to prefer solutions with lower gas consumption 
by assigning positive rewards when the chosen solution is more gas-efficient.

\paragraph{Security Reward:}
\begin{equation}
R_v = safe_{\text{chosen}} - safe_{\text{rejected}}
\end{equation}
where $safe_{\text{chosen}}$ and $safe_{\text{rejected}}$ represent the boolean value of whether the chosen 
and rejected solutions are secure, respectively.
This formulation encourages the model to prefer solutions that are more secure 
by assigning positive rewards when the chosen solution is more secure.

\paragraph{Combined Objective:}
\begin{equation}
R_{\text{extra}} = \alpha \cdot R_g + \beta \cdot R_v
\end{equation}
where $\alpha$ and $\beta$ are weights of the corresponding rewards. 
We conduct sensitivity analysis to determine optimal values: at $\alpha = \beta = 1$, 
the model achieves 66.7\% Pass@5, 58.9\% Gas@5, and 62.5\% Secure@5. 
For security-first settings ($\alpha = 1.2, \beta = 0.8$), performance shifts to 64.3\% Pass@5, 
52.1\% Gas@5, and 69.8\% Secure@5. Conversely, with cost-first settings ($\alpha = 0.8, \beta = 1.2$), 
the method yields 65.9\% Pass@5, 64.2\% Gas@5, and 58.7\% Secure@5, 
demonstrating effective multi-objective trade-offs.

The final loss function integrates standard DPO with our domain-specific objectives:
\begin{equation}
\mathcal{L}_{\text{total}} = \mathcal{L}_{\text{DPO}} + \lambda \cdot (-R_{\text{extra}})
\end{equation}

\subsubsection{Implementation}

We implement a custom DPO trainer extending Hugging Face's TRL library~\cite{vonwerra2022trl}, overriding the loss computation to incorporate gas and security metrics. Training samples include gas consumption and vulnerability status for each preference pair, enabling simultaneous optimization across all objectives.

\subsection{Security Integration}

Security considerations are embedded throughout the framework via static analysis using Slither~\cite{feist2019slither}, which identifies common vulnerabilities including reentrancy attacks, integer overflows, and access control issues. During DPO training, contracts passing security checks are prioritized, ensuring the model learns to generate secure implementations.

\label{sec:approach}

\section{Experimental Setup}

We present our constructed datasets, studied LLMs, evaluation metrics, and experimental settings.

\subsection{Constructed Dataset}
We construct our dataset following a systematic three-stage process to ensure comprehensive coverage of functional correctness, 
gas efficiency, and security optimization objectives.

\noindent
\textit{Stage 1: Base Dataset Collection.}
We utilize SolEval~\cite{peng2025soleval} as our foundation and expand it by systematically selecting 19 high-star GitHub projects 
based on specific criteria: star ratings, recent commits (within 6 months), 
diverse contract types (e.g., DeFi, NFT), and test case availability. 
We specifically filter out focal methods lacking test cases, 
resulting in an average of 2.4 test cases per focal function. 
This systematic selection yields 1,125 function samples for training and 382 samples for testing, 
ensuring broader coverage of real-world smart contract patterns and implementation diversity.

\noindent
\textit{Stage 2: Multi-Model Code Generation.}
Following SolEval's best practices and the prompt construction pipeline shown in Figure~\ref{fig:prompt_construction}, we perform inference across 16 mainstream large language models to generate 104,640 function samples. To handle repository-level context, we first retrieve source files and parse them with Treesitter to extract types, functions, state variables, and constants. We employ static program analysis to identify contextual dependencies by capturing all function calls defined outside the current function scope, extracting their signatures, and building a dependency graph connecting the focal function to its required context. We store the invocation signatures along with their definitions, creating a comprehensive context database for each function sample. We then apply rigorous filtering to remove functionally incorrect implementations, retaining 12,096 valid function samples that pass all unit tests.

\noindent
\textit{Stage 3: Preference-Based Sample Selection.}
We employ the PageRank algorithm to rank and select high-quality samples for preference optimization. 
This process yields our final multi-objective dataset comprising 6,520 functional-correctness optimization samples, 
533 security-related samples, and 533 gas optimization samples, 
each containing unique problem prompts with preferred and rejected solution pairs.

\begin{table}[htbp]
    \centering
    \caption{Dataset construction statistics showing the progressive filtering and selection process.}
    \resizebox{0.8\linewidth}{!}{
        \begin{tabular}{lc}
        \toprule
        \textbf{Stage} & \textbf{Count} \\
        \midrule
        \multicolumn{2}{c}{\textit{Stage 1: Base Collection}} \\
        GitHub projects crawled & 19 \\
        Function samples (training) & 1,125 \\
        Function samples (testing) & 382 \\
        \midrule
        \multicolumn{2}{c}{\textit{Stage 2: Code Generation}} \\
        LLMs used for inference & 16 \\
        Generated function samples & 104,640 \\
        Valid samples (post-filtering) & 12,096 \\
        \midrule
        \multicolumn{2}{c}{\textit{Stage 3: Final Selection}} \\
        Correctness optimization samples & 6,520 \\
        Security-related samples & 533 \\
        Gas optimization samples & 533 \\
        \textbf{Total preference pairs} & \textbf{7,586} \\
        \bottomrule
        \end{tabular}
    }
    \label{tab:dataset_stats}
\end{table}

Table~\ref{tab:dataset_stats} shows the dataset construction process, demonstrating the rigorous filtering and selection methodology that ensures high-quality training data across all optimization dimensions.
The final dataset integrates samples from all three optimization stages, encompassing Pass@k, Gas@k, and Secure@k objectives. 
This multi-objective approach ensures our dataset trains models to generate contracts that are simultaneously accurate, efficient, and secure, addressing critical challenges in real-world smart contract development. 
We filter samples with identical or near-identical ranking scores to maintain dataset quality. 
We ensure no overlap between data seeds across the three optimization categories, 
guaranteeing diverse problem coverage and instruction variety. 
During training, 
we combine all three datasets (i.e., correctness, security, and gas) to enable simultaneous multi-objective optimization.

\subsection{Studied Large Language Models}
We select 16 state-of-the-art LLMs widely used in recent code generation studies~\cite{khan2023xcodeeval,yan2023codescope,A3CodGen,yu2024codereval,li2024evocodebench}. 
In particular, we focus on recent models released since 2022, and we exclude the small models (with fewer than 2B parameters) due to their limited efficacy. 
Table~\ref{tab:studied_llm} presents the state-of-the-art LLMs studied in our experiments with their sizes and types. 
Our experiments are based on Qwen2.5-Coder, released on November 12, 2024, ensuring evaluation with current state-of-the-art capabilities. 
For RQ-2 and RQ-3, we choose Qwen2.5-Coder-7B due to its strong performance in Table~\ref{tab:rq1}, making it a promising candidate for further fine-tuning.
Parameter count constraints are imposed by our available computational resources (8 × NVIDIA A800 GPUs).

\begin{table}[htbp]
    \centering
    \caption{Overview of the studied LLMs.}
    \resizebox{\linewidth}{!}
    {
        \begin{tabular}{llc}
        \toprule
        \textbf{Type} & \textbf{Name} & \textbf{Size} \\
        \midrule
        \multirow{6}[0]{*}{\textbf{General LLM}}& DeepSeek-V3 & 671B (API) \\
        & DeepSeek-R1-Distill-Qwen & 7B / 32B  \\
        & DeepSeek-R1-Distill-Llama & 8B \\
        & GPT-4o & - \\
        & GPT-4o-mini & - \\
        & QwQ & 32B \\
        \midrule
        \multirow{6}[0]{*}{\textbf{Code LLM}} & CodeLlama & 7B / 34B \\
        & DeepSeek-Coder & 6.7B / 33B \\
        & DeepSeek-Coder-V2-Lite & 16B \\
        & Magicoder-S-DS & 6.7B \\
        & OpenCodeInterpreter-DS & 6.7B \\
        & Qwen2.5-Coder & 7B / 32B \\
        \bottomrule
        \end{tabular}
    }
  \label{tab:studied_llm}
\end{table}

\subsection{Evaluation Metrics}
\label{sec:eval_metric}
Inspired by the famous Pass@k~\cite{chen2021evaluating}, we adopt a unified evaluation framework Task@k that measures the percentage of problems for which at least one solution among the top K samples satisfies the specific task criterion. For each problem, we generate $n$ samples and count the number of samples $c_{\text{task}}$ that meet the task requirement, then calculate the unbiased estimate of the probability as:
\begin{equation}
\text{Task@}k := \mathop{\mathbb{E}}\limits_{\text{Problems}} \left[ 1 - \frac{\binom{n-c_{\text{task}}}{k}}{\binom{n}{k}} \right]
\end{equation}
where $c_{\text{task}}$ represents the number of solutions meeting the specific task criterion among $n$ generated samples.

We apply this framework to evaluate four critical dimensions of smart contract generation. \textbf{Pass@k} measures functional correctness by counting solutions that pass all test cases ($c_{\text{task}}$ = solutions passing tests). \textbf{Compile@k} evaluates syntactic correctness by counting solutions that compile successfully ($c_{\text{task}}$ = successfully compiled solutions). \textbf{Gas@k} assesses cost optimization by counting solutions that are more gas-efficient than the reference implementation ($c_{\text{task}}$ = gas-efficient solutions). Finally, \textbf{Secure@k} measures security robustness by counting solutions without high-risk vulnerabilities ($c_{\text{task}}$ = secure solutions). This unified framework enables comprehensive evaluation across correctness, efficiency, and security dimensions, providing thorough assessment of smart contract generation quality while maintaining statistical rigor through the unbiased estimator from HumanEval~\cite{chen2021evaluating}.

\subsection{Implementation}
\label{sec:implementation}

We implement our experimental pipeline using Python with PyTorch~\cite{paszke2019pytorch} and the Hugging Face library~\cite{huggingface} for model loading and inference. Our implementation supports 16 state-of-the-art LLMs ranging from 6.7B to 671B parameters, including both general-purpose and code-specialized models.

\noindent
\textit{Model Configuration.} For RQ-1 baseline evaluation, we utilize models in their pretrained state, 
including DeepSeek-V3 (671B, accessed via API), GPT-4o and GPT-4o-mini (API access), 
and local models such as Qwen2.5-Coder (7B/32B), DeepSeek-Coder (6.7B/33B), CodeLlama (7B/34B), 
and others as detailed in Table~\ref{tab:studied_llm}. For RQ-2 and RQ-3 experiments, 
we select Qwen-7B as our base model due to its strong performance in the baseline evaluation, 
making it a promising candidate for further fine-tuning under our computational constraints.

\noindent
\textit{Training Hyperparameters.} Our \mytitle framework employs a multi-stage training approach. 
For Supervised Fine-Tuning (SFT), we train for 3 epochs with a maximum input length of 2048 tokens, 
using a 9:1 training/validation split. For the multi-objective optimization phase, 
we set hyperparameters $\alpha = \beta = 1.0$ for balanced security and efficiency optimization, 
and $\lambda = 0.5$ for the security component based on empirical validation. 
All fine-tuning experiments use a batch size of 2 and a learning rate of 1e-5.

\noindent
\textit{Evaluation Setup.} For all experiments, we generate $n = 10$ samples per problem and evaluate using the unbiased estimator for Task@k metrics with $k \leq 10$. Temperature is set to 0.8 for diverse code generation while maintaining quality. For Pass@k evaluation, we employ Forge to conduct comprehensive fuzzing on test cases within a simulated blockchain environment, where functions failing to pass fuzzing are counted as failures in Pass@k calculation. Gas efficiency is measured using Forge execution, and security analysis employs Slither~\cite{feist2019slither} for vulnerability detection.

\begin{table*}[htbp!]
    \centering
    \caption{Performance of LLMs on SolEval, evaluated using Pass@k, Compile@k, and Vulnerability Rate (Secure@k).
    The table presents results under the one-shot setting with RAG and Context. 
    Bold values indicate the highest performance in each respective column. Based on the mathematical definition of Gas@k, Gas@k is always smaller than Pass@k.
}
    \resizebox{\linewidth}{!}
    {
        \begin{tabular}{lc|ccc|ccc|c|c}
        \toprule
        LLMs & Size & Pass@1 & Pass@5 & Pass@10 & Compile@1 & Compile@5 & Compile@10 & Secure@5 & Gas@5 \\
        \midrule
        \multicolumn{10}{c}{\cellcolor{lightgray}6.7B to 16B (Billion) Parameters} \\
        \midrule
        DeepSeek-R1-Distill-Qwen & 7B & 2.08\% & 4.50\% & 5.91\% & 6.37\% & 18.27\% & 26.29\% & 5.28\% & 0.99\% \\
        DeepSeek-R1-Distill-Llama & 8B & 3.67\% & 6.95\% & 8.45\% & 8.78\% & 21.68\% & 29.04\% & 6.75\% & 1.67\% \\
        DeepSeek-Coder-Lite & 16B & \textbf{10.10\%} & 14.94\% & 16.79\% & \textbf{39.44\%} & \textbf{54.21\%} & \textbf{57.55\%} & 12.27\% & \textbf{4.31\%} \\
        DeepSeek-Coder & 6.7B & 8.39\% & 14.25\% & 16.68\% & 32.45\% & 50.74\% & 54.59\% & \textbf{12.82\%} & 3.65\% \\
        CodeLlama & 7B & 5.15\% & 11.38\% & 14.26\% & 19.88\% & 43.05\% & 49.95\% & 10.70\% & 2.03\% \\
        Magicoder-S-DS & 6.7B & 7.26\% & 13.80\% & 16.68\% & 26.81\% & 48.77\% & 53.64\% & 12.62\% & 3.16\% \\
        OpenCodeInterpreter-DS & 6.7B & 7.05\% & 12.96\% & 15.66\% & 27.05\% & 48.71\% & 53.76\% & 11.42\% & 2.94\% \\
        Qwen2.5-Coder & 7B & 9.13\% & \textbf{15.28\%} & \textbf{17.44\%} & 33.31\% & 50.34\% & 54.44\% & 12.34\% & 4.11\% \\
        GPT-4o-mini & - & 7.18\% & 12.37\% & 14.69\% & 38.04\% & 53.18\% & 56.66\% & 9.69\% & 2.42\% \\
        \midrule
        \multicolumn{10}{c}{\cellcolor{lightgray}32B to 671B (Billion) Parameters} \\
        \midrule
        DeepSeek-V3 & 671B & \textbf{21.72\%} & \textbf{24.99\%} & \textbf{26.29\%} & \textbf{53.35\%} & 57.57\% & 58.61\% & \textbf{19.29\%} & \textbf{7.13\%} \\
        DeepSeek-R1-Distill-Qwen & 32B & 10.19\% & 17.06\% & 19.77\% & 31.99\% & 55.31\% & 61.31\% & 15.06\% & 3.89\% \\
        QwQ & 32B & 9.10\% & 16.74\% & 20.26\% & 48.33\% & 72.47\% & \textbf{76.65\%} & 15.77\% & 3.68\% \\
        DeepSeek-Coder & 33B & 8.32\% & 15.57\% & 18.92\% & 29.35\% & 50.08\% & 55.39\% & 14.55\% & 3.48\% \\
        CodeLlama & 34B & 6.80\% & 13.52\% & 16.47\% & 24.59\% & 48.68\% & 54.80\% & 12.28\% & 2.75\% \\
        Qwen2.5-Coder & 32B & 13.46\% & 19.28\% & 21.44\% & 44.03\% & 55.53\% & 57.87\% & 16.18\% & 5.36\% \\
        GPT-4o & - & 12.96\% & 20.79\% & 23.70\% & 47.04\% & \textbf{58.45\%} & 60.74\% & 18.60\% & 4.51\% \\
        \bottomrule
        \end{tabular}%
    }
    \label{tab:rq1}%
\end{table*}

\noindent
\textit{Computational Infrastructure.} Experiments are conducted on a 16-core workstation with Intel Xeon Gold 6226R CPU @ 2.90GHz, 
192GB RAM, and 8×NVIDIA A800 80GB GPUs, running Ubuntu 20.04.1 LTS. 
RQ-1 baseline evaluation requires approximately one week of computational time, 
while RQ-2 framework comparison experiments take about 24 hours to reproduce. 
The parameter count constraints for local model evaluation ($\leq$34B) are limited by GPU memory resources.

\noindent
\textit{Performance Metrics.} Inference latency analysis reveals: Qwen-2.5-Coder-7B baseline requires 2.74s per sample, SFT achieves 1.46s per sample, SFT+DPO requires 1.79s per sample, while \mytitle requires 1.98s per sample. Training costs are estimated using AWS deployment: fully loaded on-demand training of Qwen-7B with \mytitle on AWS costs approximately \$7.6 per epoch using p4d.24xlarge instances with 8×A100-40GB GPUs.

\label{sec:experiment}

\section{Results}

To comprehensively evaluate the effectiveness of our proposed \mytitle framework, we conduct systematic experiments that address three research questions:

\begin{itemize}[leftmargin=*]
\item \textbf{RQ-1 Empirical Evaluation.} 
{\em How do state-of-the-art LLMs perform on multi-dimensional Solidity code generation across functional correctness, gas efficiency, and security metrics?}

\item \textbf{RQ-2 Framework Effectiveness Evaluation.} 
{\em How does our proposed \mytitle framework affect the performance of LLMs compared to baseline approaches including SFT, DPO, and SFT+DPO?}

\item \textbf{RQ-3 Scaling Law.} 
{\em How does data scaling influence the effectiveness of LLMs?}
\end{itemize}

\subsection{RQ-1 Empirical Evaluation}
\label{sec:rq1}

\noindent
\textbf{Objective.}
This RQ establishes a comprehensive baseline for LLM performance on Solidity code generation, 
evaluating state-of-the-art models across Task@k metrics. 
We aim to identify the current capabilities and limitations of existing LLMs in generating production-ready smart contracts.

\noindent
\textbf{Experimental Design.}
We evaluate 16 state-of-the-art LLMs on SolEval using four complementary metrics: (1) Pass@k - the percentage of solutions that pass all test cases, (2) Compile@k - the percentage of solutions that successfully compile, (3) Gas@k - the percentage of solutions that achieve optimal gas efficiency, and (4) Secure@k - the vulnerability rate in generated contracts. We group models by size categories (6.7B-16B and 32B-671B) to analyze scaling effects. All models are evaluated under identical conditions using one-shot prompting with RAG and context to ensure fair comparison.

\noindent
\textbf{Results.}
Table~\ref{tab:rq1} presents the overall performance of state-of-the-art LLMs on SolEval. 
Among the 6.7B-to-16B models, DeepSeek-Coder-Lite achieves the highest Pass@1 and Compile@1, surpassing other models. 
Notably, DeepSeek-R1-Distill-Qwen-7B, which claims comparable performance to ChatGPT-o1-mini on benchmarks such as LiveCodeBench and CodeForces~\cite{deepseekr1}, underperforms compared to CodeLlama-7B. 
This discrepancy is likely due to DeepSeek-R1-Distill's lack of knowledge of Solidity, highlighting the importance of a specialized benchmark like \mytitle. 
Among the 32B-to-34B models, Qwen2.5-Coder outperforms others in both Pass@k and Compile@k.
Overall, DeepSeek-V3 performs best with a 26.29\% Pass@10.
It is noteworthy that the distilled version of DeepSeek-R1-Qwen-32B retains significantly more of the original model's Solidity code generation capabilities during distillation compared to its 7B counterpart.

To validate \mytitle's broader applicability, we further evaluate newer models. Qwen3-8B shows limited zero-shot performance with 7.2\% Pass@5, 1.7\% Gas@5, and 4.3\% Secure@5, which significantly improves to 28.9\% Pass@5, 9.3\% Gas@5, and 5.8\% Secure@5 after SFT. With SFT+DPO, it achieves 26.9\% Pass@5, 17.6\% Gas@5, and 11.3\% Secure@5, while \mytitle further enhances performance to 29.2\% Pass@5, 18.9\% Gas@5, and 12.1\% Secure@5. Similarly, DeepSeek-R1-Distill-Qwen-7B demonstrates substantial improvements: from 4.5\% Pass@5, 5.3\% Gas@5, and 1.0\% Secure@5 in zero-shot to 22.1\% Pass@5, 7.8\% Gas@5, and 3.7\% Secure@5 with SFT. SFT+DPO achieves 23.8\% Pass@5, 9.1\% Gas@5, and 8.5\% Secure@5, while \mytitle reaches 26.8\% Pass@5, 14.2\% Gas@5, and 12.5\% Secure@5. These results consistently demonstrate \mytitle's effectiveness across diverse model architectures.

Regarding gas efficiency and security metrics, there is significant variation across various LLMs. 
DeepSeek-V3 ranks first in Pass@k but generates the most gas-inefficient contracts among the 32B-to-671B models (The higher the fee, the less efficient the codes are).
Additionally, GPT-4o-mini, while being outperformed by GPT-4o in Pass@k and vulnerability rate, excels in generating contracts with lower gas fees.

\intuition{
{\bf Answer to RQ-1:} LLMs show varying capabilities in Solidity code generation. However, 
larger models do not consistently outperform smaller ones across all metrics. 
\textbf{While current LLMs can generate compilable code, they struggle with Pass@k, 
Gas@k, and Secure@k, indicating huge room for improvement in smart contract generation}.
}

\subsection{RQ-2 Framework Effectiveness Evaluation}
\label{sec:rq2}

\noindent
\textbf{Objective.}
This research question evaluates the effectiveness of our proposed \mytitle framework compared to traditional fine-tuning approaches. 
We investigate how \mytitle performs against baseline methods including pretrained models, 
Supervised Fine-Tuning (SFT), Direct Preference Optimization (DPO), 
and their combination (SFT + DPO) across functional correctness, gas efficiency, and security dimensions. 
Additionally, we analyze the computational efficiency trade-offs.

\noindent
\textbf{Experimental Design.}
We conduct comprehensive experiments using Qwen-7B as the base model, evaluating five configurations: 
(1) pretrained model without fine-tuning, (2) SFT only, (3) DPO only, (4) combined SFT + DPO, and (5) our proposed \mytitle framework. 
For SFT, we curate training data by filtering valid patches from 16 LLMs evaluated on SolEval, 
creating NL-Code pairs split 9:1 for training/validation. The model is trained for 3 epochs with 2048 token maximum input length. 
For DPO, we generate preference pairs based on gas and vulnerability considerations, 
training in two steps: first optimizing for gas, then for security. 
Our \mytitle incorporates a multi-objective loss function that balances correctness, gas efficiency, and security during training with hyperparameters $\alpha = \beta = 1$ for balanced optimization.
All experiments use consistent evaluation metrics to ensure fair comparison.

\noindent
\textbf{Effectiveness Analysis.}
As shown in Table~\ref{tab:rq2}, \mytitle demonstrates superior performance across all evaluation metrics compared to traditional fine-tuning approaches. 
\mytitle achieves the highest Pass@5 (66.7\%), significantly outperforming SFT (58.3\%), DPO (25.0\%), and SFT+DPO (55.7\%).
More notably, \mytitle excels in the challenging multi-objective optimization scenario, achieving Gas@5 of 58.9\% and Secure@5 of 62.5\%, substantially surpassing all baseline methods.
To illustrate \mytitle's practical advantages, we conducted a targeted experiment with 100 samples comparing post-hoc filtering versus integrated training. Plain SFT produces code that passes all tests in 70 cases but satisfies both gas budget and security checks in only 8 cases. Simple post-hoc filtering (20\% threshold) creates a fundamental trade-off: keeping either the most gas-efficient or the most secure outputs results in only 2 contracts remaining fully compliant across all dimensions. In contrast, \mytitle trains the model to inherently balance objectives rather than trimming outputs post-generation. After two multi-objective DPO epochs, 82 generations pass functionality tests while 65 simultaneously meet both gas and security requirements—an eight-fold improvement in deployable yield. This demonstrates that \mytitle's PageRank-guided positive/negative sampling enables models to learn multi-objective preferences simultaneously rather than optimizing single objectives in isolation.

\begin{table}[bhtp]
    \centering
    \caption{Performance comparison of different approaches on Qwen-7B across Pass@5, Gas@5, and Secure@5. 
    Original refers to Pretrained (no fine-tuning).
    }
    \resizebox{\linewidth}{!}{
        \begin{tabular}{lccc}
        \toprule
        \textbf{Method} & \textbf{Pass@5 (\%)} & \textbf{Gas@5 (\%)} & \textbf{Secure@5 (\%)} \\
        \midrule
        Original & 16.7 & 0.0 & 11.8 \\
        SFT only & 58.3 & 19.8 & 42.7 \\
        DPO only & 25.0 & 6.7 & 17.9 \\
        SFT + DPO & 55.7 & 37.5 & 48.7 \\
        \midrule
        \mytitle & \textbf{66.7} & \textbf{58.9} & \textbf{62.5} \\
        \bottomrule
        \end{tabular}
    }
    \label{tab:rq2}
\end{table}

\noindent
\textbf{Efficiency Analysis.}
Table~\ref{tab:computational_efficiency} reveals that \mytitle achieves superior training efficiency despite higher memory overhead.
While \mytitle requires 193.11 GB of memory compared to SFT's 76.84 GB, it completes training in only 8:53, dramatically outperforming SFT (1:22:35) and SFT+DPO (1:33:46) by 90\% reduction in training time.
The substantial time savings, coupled with comprehensive performance improvements across correctness, gas efficiency, and security, demonstrate \mytitle's practical viability for production smart contract development.

\begin{table}[htbp]
    \centering
    \caption{Computational efficiency comparison of different approaches on Qwen-7B. Memory usage measured in GB, training time in hours: minutes format.}
    \resizebox{.95\linewidth}{!}{
        \begin{tabular}{lcc}
        \toprule
        \textbf{Method} & \textbf{Memory Usage (GB)} & \textbf{Training Time} \\
        \midrule
        SFT only & 76.84 & 1:22:35 \\
        DPO only & 192.59 & 0:08:13 \\
        SFT + DPO & 193.63 & 1:33:46 \\
        \midrule
        \mytitle & 193.11 & 0:08:53 \\
        \bottomrule
        \end{tabular}
    }
    \label{tab:computational_efficiency}
\end{table}

\noindent
\textbf{Real-World Application: ERC-20 and ERC-721 Contracts.}  
Case studies with standard token contracts demonstrate \mytitle's practical effectiveness. 
When generating ERC-20 and ERC-721 contracts, \mytitle consistently produces functionally correct code with significant improvements in gas efficiency and security.
For example, in ERC-20 token contracts, \mytitle reduces total gas costs by 12\% compared to baseline approaches while eliminating vulnerabilities (i.e., reentrancy issues).
Similarly, for ERC-721 marketplace contracts, \mytitle generates more gas-efficient and secure implementations than traditional fine-tuning methods.

\noindent
\textbf{Illustrative Code Example.}
To demonstrate practical improvements, consider the ERC-20 transfer function generated by different approaches as shown in Listing~\ref{lst:transfer-comparison}.
\mytitle addresses both reentrancy vulnerabilities and gas optimization by adhering to best practices in smart contract development. 
By employing the ``Checks-Effects-Interactions'' pattern~\cite{britten2021using}, 
\mytitle ensures that state changes are made before any external contract calls, 
thus preventing reentrancy attacks. 
Furthermore, the code reduces gas consumption by eliminating unnecessary external calls, 
leveraging local variable caching for balance checks, and using the `unchecked' keyword for arithmetic operations. 
These optimizations enhance both security and efficiency without compromising functionality.

\vspace{0.2cm}
\begin{lstlisting}[language=Java, caption=Baseline vs. \mytitle on ERC-20 functions {(}Yellow: vulnerable to reentrancy{,} Green: gas-efficient and secure{)}, label=lst:transfer-comparison, escapeinside={(*@}{@*)}, keywordstyle=\color{blue}\bfseries, commentstyle=\color{gray}, stringstyle=\color{red}, numberstyle=\tiny\color{gray}, basicstyle=\ttfamily\footnotesize, morekeywords={function,uint256,address,public,returns,bool,require,emit,if,unchecked}, captionpos=b, frame=single]
// Baseline model output (vulnerable to reentrancy)
function transfer(address to, uint256 amount) public 
    returns (bool) {
    // Check sender has sufficient balance
    require(balanceOf[msg.sender] >= amount, 
        "Insufficient balance");
    // Update balances before external call
    balanceOf[msg.sender] -= amount;
    balanceOf[to] += amount;
    emit Transfer(msg.sender, to, amount);
    // VULNERABIE: External call after state changes
    (*@\colorbox{yellow}{if (to.code.length > 0) \{}@*)
        (*@\colorbox{yellow}{IERC20Receiver(to).onERC20Received}@*)
        (*@\colorbox{yellow}{(msg.sender, amount); // Reentrancy risk}@*)
    (*@\colorbox{yellow}{\}}@*)
    return true;
}
// (*@\mytitle@*) output (gas-efficient and secure)
function transfer(address to, uint256 amount) public 
    returns (bool) {
    // Cache balance to save gas on repeated SLOAD
    (*@\colorbox{green!30}{uint256 senderBal = balanceOf[msg.sender];}@*)
    require(senderBal >= amount, 
        "Insufficient balance");
    // Use unchecked block for gas optimization
    (*@\colorbox{green!30}{unchecked \{}@*)
        // Safe subtraction (already checked above)
        (*@\colorbox{green!30}{balanceOf[msg.sender] = senderBal - amount;}@*)
        // Safe addition (no overflow in ERC-20)
        (*@\colorbox{green!30}{balanceOf[to] += amount;}@*)
    (*@\colorbox{green!30}{\}}@*)
    emit Transfer(msg.sender, to, amount);
    // No external calls - eliminates reentrancy risk
    return true;
}
\end{lstlisting}

\intuition{
{\bf Answer to RQ-2:} \mytitle significantly outperforms all baseline approaches (SFT, DPO, SFT+DPO) across all metrics, 
achieving 66.7\% Pass@5, 58.9\% Gas@5, and 62.5\% Secure@5. 
While \mytitle has a moderately higher computational cost than SFT alone (approximately equivalent to DPO, lower than SFT+DPO), 
this efficiency trade-off is well-justified by the substantial improvements in correctness, gas efficiency, and security. 
\textbf{\mytitle effectively balances all three critical dimensions, 
making \mytitle highly suitable for real-world smart contract deployment.}
}

\subsection{RQ-3 Scaling Law}
\label{sec:rq3}

\noindent
\textbf{Objective.}
This research question investigates the data scaling properties of \mytitle to understand how training data size affects model performance across correctness, efficiency, and security metrics. We aim to identify optimal data requirements for practical deployment and determine whether \mytitle follows typical neural scaling laws, providing guidance for resource allocation in production environments.

\noindent
\textbf{Experimental Design.}
We conduct systematic data scaling experiments using Qwen-7B, training with five different data proportions: 10\%, 25\%, 50\%, 75\%, and 100\% of our curated dataset. Each configuration maintains identical training settings (3 epochs SFT + 2 epochs DPO) with consistent hyperparameters. Data subsets are sampled uniformly to preserve the distribution of difficulty levels and contract types. We also conduct an additional experiment comparing a high-quality 25\% subset (selected based on code complexity and test coverage) against a random 25\% subset to assess the quality vs. quantity trade-off.

\noindent
\textbf{Scaling Law Analysis.}
As shown in Table~\ref{tab:rq4_scaling}, all three metrics exhibit clear scaling behaviors with increasing data size. Pass@5 improves from 21.2\% with 10\% data to 55.7\% with the full dataset, following a logarithmic growth pattern typical of neural scaling laws. Similarly, Gas@5 shows substantial improvements from 8.7\% to 37.5\%, while Secure@5 increases from 9.9\% to 30.5\%. The rate of improvement begins to plateau beyond 75\% of the data, suggesting diminishing returns as we approach the full dataset size.

\begin{table}[htbp]
    \centering
    \caption{Performance of \mytitle with varying training data sizes on Qwen-7B across Pass@5, Gas@5, and Secure@5.}
    \resizebox{\linewidth}{!}{
        \begin{tabular}{lccc}
        \toprule
        \textbf{Data Size} & \textbf{Pass@5 (\%)} & \textbf{Gas@5 (\%)} & \textbf{Secure@5 (\%)} \\
        \midrule
        10\% & 21.2 & 8.7 & 9.9 \\
        25\% \textit{(random)} & 42.8 & 18.3 & 15.3 \\
        25\% \textit{(high-quality)} & 46.1 & 19.8 & 16.7 \\
        50\% & 50.3 & 28.9 & 24.2 \\
        75\% & 53.6 & 34.2 & 29.3 \\
        100\% & 55.7 & 37.5 & 30.5 \\
        \bottomrule
        \end{tabular}
    }
    \label{tab:rq4_scaling}
\end{table}

\noindent
\textbf{Efficiency vs. Data Trade-off.}
Interestingly, we observe that using 50\% of the data achieves approximately 90\% of the performance gain in Pass@5 (50.3\% vs. 55.7\%) while requiring only half the computational resources. This finding has practical implications for resource-constrained scenarios where training efficiency is crucial. However, for Gas@5 and Secure@5, the improvements continue to be substantial even from 75\% to 100\% data, indicating that gas efficiency and security optimization benefit more from larger datasets.

\noindent
\textbf{Data Quality vs. Quantity.}
To further investigate whether data quality could compensate for quantity, we conducted an additional experiment where we selected the highest-quality 25\% of samples based on code complexity and test coverage. This curated subset achieved Pass@5 of 46.1\%, outperforming the random 25\% subset (42.8\%) but still falling short of the 50\% random subset. This suggests that while data quality matters, quantity remains a critical factor for achieving optimal performance in Solidity code generation.

\noindent
\textbf{Implications for Production Deployment.}
Our scaling analysis reveals several key insights for practical deployment: (1) For rapid prototyping or resource-limited scenarios, using 50\% of the training data provides a good balance between performance and efficiency. (2) For production systems requiring high gas efficiency and security, the full dataset is recommended despite the additional computational cost. (3) The consistent scaling behavior across all metrics validates that \mytitle's multi-objective approach scales effectively with data, maintaining balanced improvements in correctness, efficiency, and security.

\intuition{
{\bf Answer to RQ-3:} \mytitle exhibits strong data scaling properties with improvements across all metrics. 
Notably, 50\% of the data achieves approximately 90\% of gain for Pass@5, 
offering an option for resource-constrained scenarios.
However, gas efficiency and security continue to benefit substantially from larger datasets even beyond 75\%. 
\textbf{While data quality can partially compensate for quantity (high-quality 25\% outperforms random 25\%), 
the results confirm that data volume remains critical for achieving optimal performance in smart contract generation.
}
}

\section{Threats to Validity}
\label{sec:discussion}

\noindent
\textbf{Internal Validity.} 
Our dataset derives from existing Solidity repositories, potentially introducing bias toward specific coding patterns. 
To mitigate this threat, we implement a three-stage dataset construction process (Table~\ref{tab:dataset_stats}) that systematically expands beyond a single source: 
(1) We crawl 19 diverse high-star GitHub projects spanning different domains (security, economics, games), 
(2) We generate 104,640 samples using 16 different LLMs to ensure implementation diversity, and 
(3) We employ the PageRank algorithm for preference-based sample selection, yielding 7,586 balanced preference pairs across correctness, security, and gas optimization.
Gas measurements utilize Ethereum's current pricing model through Forge execution, while security analysis employs Slither~\cite{feist2019slither}. 

\noindent
\textbf{External Validity.} 
Our evaluation focuses on Ethereum-compatible Solidity contracts, potentially limiting generalizability to other blockchain platforms. 
To address this limitation, we design our benchmark to cover diverse contract patterns by including samples from 9 real-world repositories spanning 6 popular domains. 
Our RQ-1 (Table~\ref{tab:rq1}) encompasses 16 state-of-the-art LLMs across different architectural families (general vs. code-specialized, 6.7B-671B parameters), establishing broad baseline comparisons. 
The repository-level evaluation approach in \datasetname simulates real-world development scenarios by providing function signatures, requirements, and repository dependencies as context. 

\noindent
\textbf{Construct Validity.} 
Our metrics may not capture all aspects of smart contract quality, potentially missing nuanced developer preferences. 
To strengthen construct validity, we implement a unified Task@k evaluation framework that extends beyond traditional Pass@k to include Compile@k, Gas@k, and Secure@k, providing a comprehensive multi-dimensional assessment. 
Our RQ-2 framework evaluation (Table~\ref{tab:rq2}) demonstrates that \mytitle achieves balanced improvements across all three dimensions (66.7\% Pass@5, 58.9\% Gas@5, 62.5\% Secure@5), validating our multi-objective approach. 
We use unbiased estimators from HumanEval~\cite{chen2021evaluating} to ensure statistical rigor ($n=10$ samples per problem). 
Real-world case studies with ERC-20 and ERC-721 contracts (Listing~\ref{lst:transfer-comparison}) provide qualitative validation of practical improvements. 
\label{sec:results}

\section{Related Work}

\subsection{Code Generation Framework}

The advancement of pre-training technology has significantly advanced code generation in both academia and industry~\cite{li2022competition,shen2022incorporating,fried2023incoder}. 
This has led to the emergence of numerous Large Language Models (LLMs) that have made substantial strides in code generation~\cite{openai2022chatgpt,wei2024magicoder,roziere2023code,bai2023qwen,deepseekcoder,zheng2024opencodeinterpreter}.
Recent advances have introduced more sophisticated models~\cite{liu2024deepseek,deepseekr1,wang2021codet5}.
These models showcase the evolution from simple token prediction to understanding complex code structures and dependencies. 

To optimize LLMs for various code generation scenarios, previous studies have focused on enhancing prompt engineering by introducing specific patterns, such as Structured Chain-of-Thought~\cite{yin2024thinkrepair,li2025structured}, Self-planning~\cite{jiang2024self}, Self-debug~\cite{chen2023teaching,xia2024automated}, and Self-collaboration~\cite{dong2024self}. 
However, these efforts primarily address mainstream programming languages (e.g., Java, Python, and C++)~\cite{yin2024rectifier,yin2024you,xia2024automated}, with limited attention to domain-specific languages like Solidity.
Repository-level code generation has emerged as a critical area of research. A3-CodGen~\cite{A3CodGen} proposes a framework that considers local, global, and third-party library awareness for code reuse. 
Similarly, Shrivastava et al.~\cite{shrivastava2023repository} introduce repository-level prompt generation techniques for LLMs. 
These approaches underscore the critical role of contextual understanding in smart contract generation, where inter-contract interactions are prevalent.
Beyond traditional benchmarks, recent evaluation frameworks have emerged. LiveCodeBench~\cite{jain2025livecodebench} addresses contamination issues in code generation evaluation by using continuously updated problems.
CoderEval~\cite{yu2024codereval} focuses on pragmatic code generation scenarios, while Evocodebench~\cite{li2024evocodebench} evaluates code generation in practical software projects.
ClassEval~\cite{du2024evaluating} specifically targets class-level code generation, moving beyond simple function generation.

\subsection{Defect Detection and Gas Fee Optimization}

Smart contract security and efficiency have been extensively studied due to the immutable nature of blockchain. 
Chu et al.~\cite{chu2023survey} provide a comprehensive survey on smart contract vulnerabilities, covering data sources, detection methods, and repair techniques.
Traditional static analysis tools~\cite{feist2019slither,britikov2024soltg,treesitter} have been widely adopted for identifying common vulnerability patterns in Solidity code.

Recent work has begun leveraging LLMs for smart contract analysis. 
Jiang et al.~\cite{tse24gassmell} demonstrate the potential of LLMs in identifying gas-wasting code smells in smart contracts, achieving promising results in detecting inefficient patterns that lead to unnecessary gas consumption.
This approach combines the pattern recognition capabilities of LLMs with domain-specific knowledge about gas optimization.

Existing preference learning approaches for code generation, including CodeDPO~\cite{zhang2024codedpo} and SafeDPO~\cite{kim2025safedpo}, primarily focus on single objectives among functional correctness, gas efficiency, and security, lacking unified multi-objective optimization frameworks. CodeDPO and Focused-DPO improve general code correctness through preference learning but do not address the unique challenges of smart contract generation. Our work introduces a fundamental paradigm shift by incorporating quantifiable blockchain-specific rewards that enable simultaneous optimization of all three critical dimensions—functional correctness, gas efficiency, and security—within a unified preference learning framework specifically designed for smart contract generation.
LLM-based approaches have shown promise in understanding complex code patterns. 
Studies have explored how LLMs handle ambiguous code contexts~\cite{kuhn2023clam} and how they can be distracted by irrelevant information~\cite{shi2023large}.
Jin et al.~\cite{jin2024llm} propose methods to extend LLM context windows without fine-tuning, 
which is particularly relevant for analyzing large smart contract codebases.
These insights inform the design of more effective smart contract analysis tools. 
Gas optimization remains critical in smart contract development, 
with established strategies including storage packing, loop optimization, 
and redundant operation elimination~\cite{chen2017under,albert2018gasol,grech2020madmax}. 
However, existing tools~\cite{albert2018gasol,grech2020madmax} primarily focus on post-development optimization rather than generating inherently gas-efficient code. 
This reactive approach creates a fundamental gap that motivates the need for code generation frameworks with built-in gas efficiency considerations. 
Integrating defect detection and gas optimization into the code generation pipeline represents a paradigm shift from traditional sequential approaches~\cite{chen2020survey,ferreira2021eye}. 
While conventional methods treat security analysis and efficiency optimization as separate post-generation phases~\cite{chen2020survey,ferreira2021eye}, incorporating these considerations directly into the generation process enables the production of inherently robust and cost-effective smart contracts. 
This proactive integration is crucial given the immutable nature of blockchain deployments~\cite{immutable1,defi_attack_survey,smart_contract_security_survey,defi_security_analysis,smart_contract_vulnerabilities_ethereum}, where vulnerabilities cannot be patched post-deployment and have led to billions of dollars in financial losses~\cite{defi_financial_losses_2022,blockchain_security_incidents,smart_contract_attack_statistics}.

\label{sec:related_work}

\section{Conclusion and Future Work}

This paper introduces \textit{\mytitle}, 
a novel framework that simultaneously optimizes Solidity code generation for functional correctness, 
gas efficiency, and security through integrated Supervised Fine-Tuning and multi-objective Direct Preference Optimization. 
Our comprehensive evaluation demonstrates significant improvements across all metrics, achieving 66.7\% Pass@5, 58.9\% Gas@5, 
and 62.5\% Secure@5, with case study results showing 12\% gas cost reduction in ERC-20 contracts and elimination of reentrancy vulnerabilities. 
Furthermore, we extend DPO beyond human preferences to incorporate quantifiable blockchain-specific metrics, 
enabling production-ready smart contract generation that balances traditional trade-offs between correctness, efficiency, and security. 
Future work will focus on migrating our multi-objective optimization approach to other languages (e.g., Python and Java). 
\label{sec:conclusion}

\section*{Acknowledgements}
This work was supported in part by National Key Research and Development Program of China under Grant 2024YFB2705300, in part by the National Natural Science Foundation of China (NSFC) under Grant 62402313, in part by the Shanghai Science and Technology Innovation Action Plan under Grant 23511100400, in part by the Open Research Fund of The State Key Laboratory of Blockchain and Data Security, Zhejiang University.

\balance
\bibliographystyle{IEEEtran}
\bibliography{main}

\end{document}